\documentclass[aps,prl,twocolumn,groupedaddress]{revtex4}
\usepackage{epsfig}
\usepackage{amsmath}
\usepackage{amsfonts}
\def\bE{\mathbf E}

\def\bx{\mathbf x}

\def\bk{\mathbf k}

\def\b0{\mathbf 0}

\begin{document}


\title{Classification of electromagnetic resonances in finite\\ inhomogeneous three-dimensional
structures}
\author{Neil V. Budko}
\affiliation{Laboratory of Electromagnetic Research, Faculty of Electrical Engineering, Mathematics and Computer Science,
Delft University of Technology,
Mekelweg 4, 2628 CD Delft, The Netherlands}
\email{n.budko@ewi.tudelft.nl}
\author{Alexander B. Samokhin}
\affiliation{Department of Applied Mathematics, Moscow Institute of
Radio Engineering, Electronics, and Automatics (MIREA), Verndasky~av.~78, 117454, Moscow, Russian Federation}
\thanks{This research is supported by NWO (The Netherlands)
and RFBR (Russian Federation).}



\date{\today}

\begin{abstract}
We present a simple and unified classification of macroscopic electromagnetic resonances in
finite arbitrarily inhomogeneous isotropic dielectric 3D structures situated in free space. By observing the complex-plane 
dynamics of the spatial spectrum of the volume integral operator as a function of angular frequency and 
constitutive parameters we identify and generalize all the usual resonances, including complex plasmons, 
real laser resonances in media with gain, and real quasi-static resonances in media with 
negative permittivity and gain. 
\end{abstract}


\maketitle
It is hard to overestimate the role played by macroscopic electromagnetic resonances in physics.
Phenomena and technologies such as lasers, photonic band-gap materials, plasma waves
and instabilities, microwave devices, and a great deal of electronics are all related or even entirely 
based on some kind of electromagnetic resonance.
The usual way of analysis consists of deriving the so-called dispersion equation, which
relates the wave-vector $\bk$ or the propagation constant $\vert\bk\vert$ of a plane 
electromagnetic wave to the angular frequency $\omega$. The solutions of this equation may be real 
or complex. In the first case we talk about a {\it real resonance}, i.e. such that can be attained 
for some real angular frequency and therefore, in principle, results in unbounded fields. In reality,
however, amplification of the fields is bounded by other physical mechanisms, e.g. nonlinear saturation.
If solution is complex, then we have a {\it complex resonance} and, depending on the
sign of the imaginary part, the associated fields are either decaying or growing with time.
This common approach is rather limited and does not include all pertaining phenomena.
Indeed, more or less explicit dispersion equations can only be obtained for infinite 
(unbounded) homogeneous media, as often done in plasma and photonic studies.
Other approaches impose explicit boundary conditions and can handle resonators and waveguides with 
perfectly conducting walls, and idealistic piece-wise homogeneous objects (e.g. plane layered medium, 
circular cylinders, a sphere). On the other hand, very little can be said in the general case of a finite 
inhomogeneous dielectric object situated in free space. Due to the absence of an explicit dispersion equation
and explicit boundary conditions, 
even the existence and classification of resonances in such objects is still an open problem. 

We describe here an alternative mathematically rigorous approach to electromagnetic resonances, based 
on the volume integral formulation of the electromagnetic scattering, also known as
the Green's function method and the domain integral equation method.
This formulation is equivalent to the Maxwell's equations and is perfectly suited for
bounded inhomogeneous objects in free space. Despite its generality, nowadays the volume integral 
equation is mostly used as a numerical tool, for instance, in near-field optics and geophysics.
The main limitation seems to be the implicit mathematical structure of this equation resisting any straightforward 
analysis and interpretation. Recently, however, we have succeeded in deriving useful mathematical bounds on 
the spatial spectrum of the volume integral operator proving, in particular, that along with the usual discrete eigenvalues 
this operator has a dense essential spectrum as well \cite{BudkoSamokhinSIAM2005}. Below we 
reiterate our results and show how to use them in the analysis of resonances. Then, we proceed with a step by step classification
of all known complex and real resonances. In particular, we generalize the notion of a complex plasmon, real laser resonance, 
and a real quasi-static resonance in an exotic material containing a negative permittivity part and a part with gain. 
Recently, several authors have suggested \cite{Bergman2001}--\cite{Pendry2003} that this type of material may be an answer 
to some urgent technological questions ranging from surface plasmon lasers (SPASER) to loss compensation in media with 
negative refraction (perfect lens). 
We believe that our analysis provides a necessary generalization and a handy
analytical tool for these and other studies, especially in what concerns the resonant light confinement. 

The frequency-domain Maxwell's equations describing the electromagnetic field in a non-magnetic 
isotropic inhomogeneous object occupying finite spatial domain $D$ lead to the following strongly singular 
integral equation:
 \begin{align}
 \label{eq:SingularVIE}
 &\bE^{\rm in}(\bx,\omega)
 = 
 \left[{\mathbb I} + \frac{1}{3}\chi(\bx,\omega)\right]
 \bE(\bx,\omega)
 \\ \nonumber
 & - \lim\limits_{\epsilon\rightarrow 0}
 \int\limits_{\bx'\in D \setminus\vert\bx-\bx'\vert<\epsilon}
 {\mathbb G}(\bx-\bx',\omega)
 \chi(\bx',\omega)
 \bE(\bx',\omega)
 \;{\rm d}\bx',
 \end{align}
where ${\mathbb I}$ denotes a unit tensor ($3\times 3$ identity matrix), whereas the explicit form of the 
Green's tensor ${\mathbb G}$ is of no importance here, but can be found in \cite{Rahola2000} and \cite{SamokhinBook}.
Here, $\bE^{\rm in}$ is the incident field
in vacuum background, where the wavenumber is $k_{\rm 0}=\omega/c$, and the total electric field 
in the configuration is denoted by $\bE$. Constitutive parameters of the object are contained in 
the so-called contrast function $ \chi(\bx,\omega)=\varepsilon_{\rm r}(\bx,\omega)-1$,
where $\varepsilon_{\rm r}$ is the relative dielectric permittivity of the object.
In operator notation equation (\ref{eq:SingularVIE}) can be written simply as
 \begin{align}
 \label{eq:OperatorVIE}
 Au=u^{\rm in}.
 \end{align}
The spatial spectrum of operator $A$ is defined as a set $\sigma(\lambda)$ of complex numbers $\lambda$ 
for which operator
 \begin{align}
 \label{eq:Resolvent}
 [A-\lambda I]^{-1}
 \end{align}
fails to exist in one or another way. We need to distinguish here two cases. The first is when for some $\lambda$ 
the homogeneous equation $[A-\lambda I]u_{\lambda}=0$
has a nontrivial solution $u_{\lambda}\ne 0$. In addition, this solution has a finite norm, i.e. $\Vert u_{\lambda}\Vert<\infty$.
If the latter condition is satisfied, then $\lambda$
is called an {\it eigenvalue} and the corresponding $u_{\lambda}$ -- an {\it eigenfunction} 
({\it eigenmode}). It happens that eigenvalues constitute, although possibly infinite, but {\it discrete} 
subset of the complex plane -- a set of isolated points, in other words.

The second case is when equation $[A-\lambda I]u_{\lambda}=0$ is formally satisfied by some $u_{\lambda}$, which
either does not have a bounded norm, i.e. $\Vert u_{\lambda}\Vert\rightarrow\infty$, or is localized to a
single point in space. The set of $\lambda$'s
corresponding to such cases is often a dense subset of the complex plane, sometimes referred to as {\it essential} spectrum.
An even more rigorous analysis would also require distinction 
between the {\it continuous} and the {\it residual} spectra, however, so far we cannot come-up with a simple formal rule
to identify and separate them  in the electromagnetic case. It is quite easy to find the physical interpretation 
of $\Vert u_{\lambda}\Vert\rightarrow\infty$. For example, in the $L^{2}$ norm suggested by the electromagnetic 
energy considerations (Pointing's theorem), such functions are a plane wave and the Dirac's delta function, which both
have infinite $L^{2}$ norms. The essential spectrum associated with plane waves is common for infinite periodic
structures, where it surrounds photonic band gaps, and in infinite plasma models, where it gives rise to 
certain types of plasma waves.

In \cite{BudkoSamokhinSIAM2005} we prove that the strongly singular integral operator of equation (\ref{eq:SingularVIE})
has both the dense essential spectrum and the discrete eigenvalues. Moreover, for any inhomogeneous
object with $\chi(\bx,\omega)$ H{\"o}lder-continuous in ${\mathbb R}^{3}$ (i.e. inside the object as well as across 
its outer boundary) the essential spectrum $\lambda_{\rm ess}$ is given explicitly as
 \begin{align}
 \label{eq:EssentialSpectrum}
 \lambda_{\rm ess}=\varepsilon_{\rm r}(\bx,\omega), \;\;\;\; \bx\in{\mathbb R}^{3}.
 \end{align}
In other words $\lambda_{\rm ess}$ will consist of all values of $\varepsilon_{\rm r}$, which 
it admits in ${\mathbb R}^{3}$. Thus it will always contain the real unit, since it is the relative
permittivity of vacuum, and a curve or even an area of the complex plane emerging from the
real unit and running through all other values, which macroscopic $\varepsilon_{\rm r}$ takes inside the object.
This part of the spectrum does not depend on the object's size or shape, or even the relative volume occupied by
different inhomogeneities.

In addition to the essential spectrum operator $A$ has the usual discrete eigenvalues located within 
the following wedge-shaped bounds:
 \begin{align}
 \label{eq:Eigenvalues}
 \begin{split}
 &{\rm Im}\,\varepsilon_{\rm r}(\bx,\omega)\left[1-{\rm Re}\lambda\right]+
 \\
 &\left[{\rm Re}\,\varepsilon_{\rm r}(\bx,\omega)-1\right]{\rm Im}\lambda
 \le 0,
 \;\;\bx\in D.
 \end{split}
 \end{align}
It is also known that $\vert\lambda\vert\le\Vert A\Vert$, and that $\Vert A\Vert<\infty$ for any $\chi$, H{\"o}lder-continuous in ${\mathbb R}^{3}$. 
Exact distribution of eigenvalues in the complex plane is unknown to us and depends on the object's shape.
The eigenfunctions (modes) associated with these eigenvalues are global (not localized) and, in general, can only 
be found numerically. 
\begin{figure*}[t]
\epsfig{file=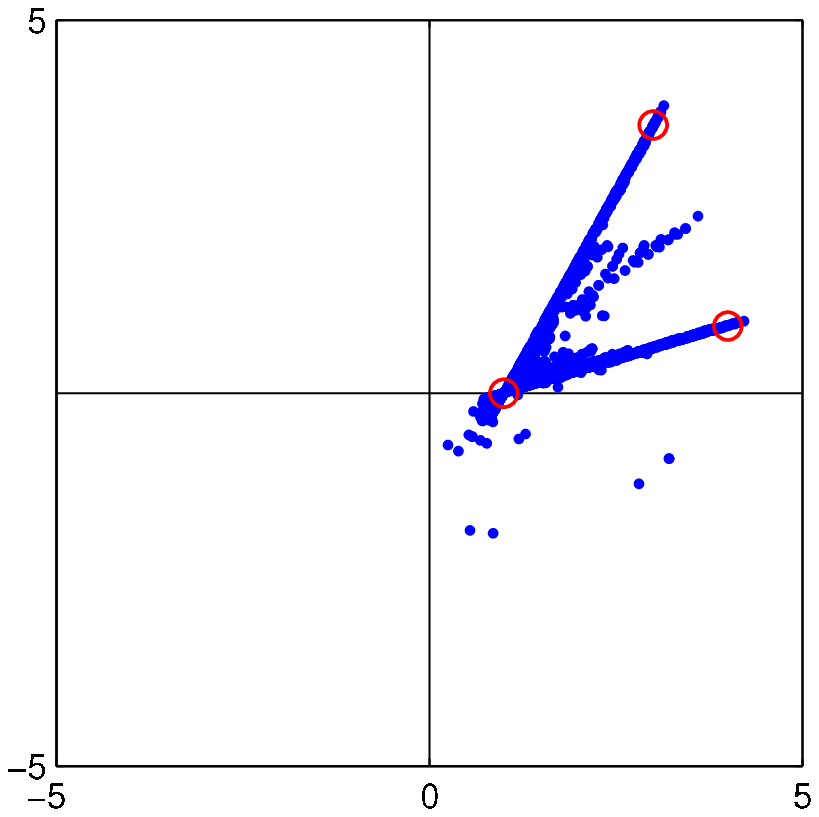,width=0.6\columnwidth,height=0.6\columnwidth}
\epsfig{file=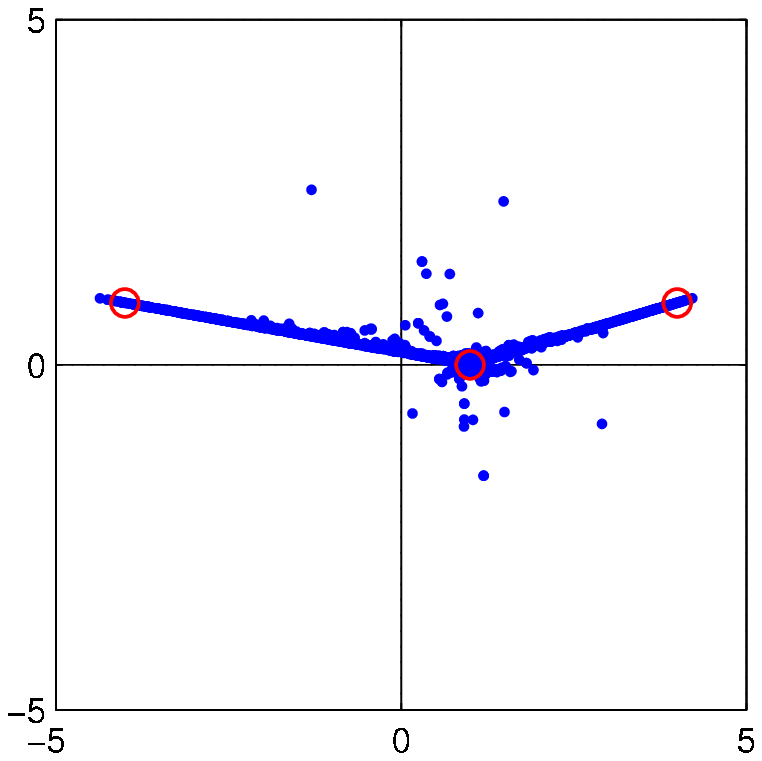,width=0.6\columnwidth,height=0.6\columnwidth}
\epsfig{file=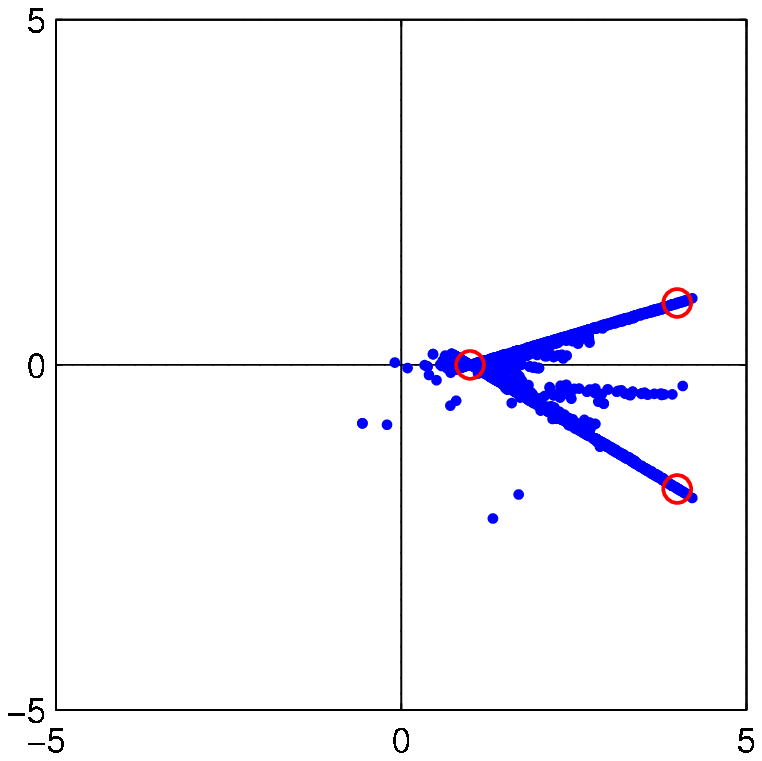,width=0.6\columnwidth,height=0.6\columnwidth}
\caption{Known resonances as seen in the numerical spectrum for various inhomogeneous objects. Left:
object with losses (complex wave-like resonances). Middle: object with strong anomalous dispersion (complex plasmon resonance).
Right: object with lossy and gain parts (real wave-like laser resonance).}
\end{figure*}

To use these results in the analysis of electromagnetic resonances we note that both the essential spectrum 
and the eigenvalues are parametric functions of the angular frequency $\omega$. 
In general, a perfect (real) resonance would occur, if for some $\omega$ the spatial spectrum of $A$ 
would acquire a zero eigenvalue. If, on the other hand, for some $\omega$ the spatial spectrum 
does not contain zero, but gets close to it, while moves away for other $\omega$'s, then we
have a complex resonance. With this in mind, one should try to visualize the dynamics of the 
spatial spectrum as it `moves' in the complex plane, paying attention to the eigenvalues and portions
of essential spectrum, which first approach zero and then move away from it. 
Expression (\ref{eq:EssentialSpectrum}) is very important in this respect as it tells us that the motion of 
essential spectrum is explicitly related to the temporal dispersion of the relative permittivity. 
We also know (see below) that the eigenfunctions related to this spectrum are highly localized. Thus
from (\ref{eq:EssentialSpectrum}) and the known spatial distribution of $\varepsilon_{\rm r}(\bx,\omega)$
we can immediately tell where exactly in $D$ would a local resonance occur.
The motion of discrete eigenvalues, on the other hand, is quite unpredictable, with the general 
tendency to spread out at higher frequencies. While doing so, some of these eigenvalues may pass
through or close to zero, which will be an indication of a global resonance. We propose here a useful 
rule of thumb for visualizing the eigenvalue bound (\ref{eq:Eigenvalues}). Imagine a line drawn through 
the real unit and any value of $\varepsilon_{\rm r}$ inside the object. If you now stand in the complex plane and 
look from the real unit towards that value of $\varepsilon_{\rm r}$, then the eigenvalues can only be to your right.

Finally, we have also been able to prove that in the static 
limit $\omega\rightarrow 0$ or $D\rightarrow 0$ {\it all} discrete
eigenvalues are located within the convex envelope of essential spectrum \cite{BudkoSamokhinDIFUR2005}, and
are given by
 \begin{align}
 \label{eq:Static}
 \lambda=\frac{\int\limits_{\bx\in D}\varepsilon_{\rm r}(\bx,\omega)\vert\nabla\varphi_{\lambda}(\bx)\vert^{2}\,{\rm d}\bx}
 {\int\limits_{\bx\in D}\vert\nabla\varphi_{\lambda}(\bx)\vert^{2}\,{\rm d}\bx},
 \end{align}
where $\varphi_{\lambda}$ is a scalar static mode.
Formally, our essential spectrum (\ref{eq:EssentialSpectrum}) can be derived from this expression as well,
by taking $\vert\nabla\varphi_{\lambda}(\bx)\vert^{2}\sim\delta(\bx-\bx')$.
This also proves that the eigenfunctions associated with the essential spectrum
are highly localized in space. Another important observation is about the
discrete eigenvalues {\it outside} the convex envelope of essential spectrum. 
Since those do not exist in the quasi-static regime and appear only at higher frequencies and object sizes, 
we may conclude that the corresponding eigenfunctions are not of static type, but more of the wave-like type, 
i.e. oscillating in space. 

Now, we have everything we need for a unified description of resonances.
We shall illustrate our conclusions by numerically computed spectra
for an inhomogeneous cube consisting of two equal halves with different permittivity values. 
The side of the cube is half of the vacuum wavelength.

In objects consisting of lossy dielectric materials only complex resonances can be observed.
For example, in Fig.~1 (left) we show the spectrum for the case of a lossy dielectric with both
${\rm Re}\,\varepsilon_{\rm r}>0$ and ${\rm Im}\,\varepsilon_{\rm r}>0$.
The actual values of relative permittivity and the real unit are given as circles. 
Numerical equivalent of essential spectrum (there is no such thing as dense or continuous spectrum with matrices) 
always looks like a set of line segments emerging from the real unit \cite{BudkoSamokhinSIAM2005}. 
One should simply keep in mind that in a continuously inhomogeneous object this spectrum may be a 
rather arbitrary curve or an area. Other, off-line eigenvalues
are within the bounds prescribed by (\ref{eq:Eigenvalues}). As the angular frequency varies,
some of these latter off-line eigenvalues may get close, but not equal to zero. These are the complex 
resonances, corresponding to complicated global wave-like spatial modes.

In Fig.~1 (middle) we illustrate the case where due to strong anomalous dispersion one of the object's parts has 
${\rm Re}\,\varepsilon_{\rm r}<0$ and ${\rm Im}\,\varepsilon_{\rm r}>0$ at a certain angular frequency.
The line of essential spectrum proceeds close to the zero of the complex plane. For other
angular frequencies this line will move away from zero. It is well-known that this combination of materials 
supports complex plasmon resonances. Hence, we may safely conclude that we deal here with one of them.
As an extra confirmation we see that this resonance is related to the highly localized modes of essential spectrum. 
Further we conclude that, in general, complex plasmons may exist not only at an interface between two homogeneous objects, 
but along rather arbitrary surfaces inside a continuously inhomogeneous object with strong anomalous dispersion. The 
precise location of this surface is determined by that value of $\varepsilon_{\rm r}$ inside $D$, 
which appears to be the closest to zero.

Recalling the rule of thumb about the location of eigenvalues we realize that a discrete eigenvalue
can be equal to zero, only if the relative permittivity at some point inside the object happens 
to have a negative imaginary part, i.e. ${\rm Im}\,\varepsilon_{\rm r}<0$. This corresponds to the so-called negative losses
or gain as in pumped laser media. In Fig.~1 (right) the numerical spectrum for a cube with one lossy half and 
another half with gain is shown. Two of the discrete eigenvalues
are very close to zero, meaning that the whole configuration is in the vicinity of a real laser resonance.
It is, however, very hard to come-up with an exact real resonance in this way. For a given temporal dispersion of the
medium, one has to optimize the geometrical parameters of the object until the resonance is achieved,
which is a very challenging numerical problem. One thing we can be sure about, though:
for such configurations the zero eigenvalue will always be {\it outside} the convex envelope of the essential 
spectrum. Therefore, real laser resonances correspond to wave-like spatial modes and, thus, can
only be achieved in structures whose size is comparable to or greater than the medium wavelength.
This is confirmed by the standard theory of lasers.

\begin{figure}[t]
\centerline{\epsfig{file=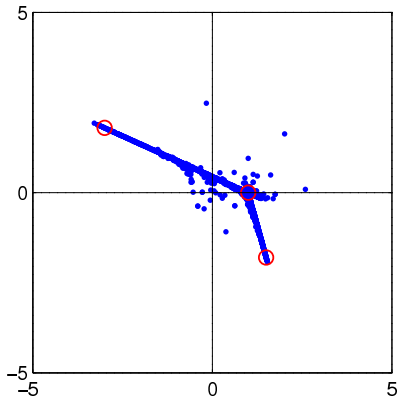,width=0.6\columnwidth,height=0.6\columnwidth}}
\caption{Real quasi-static resonance in an object with negative permittivity and gain.}
\end{figure}
As we already mentioned in the beginning, combination of a negative permittivity material and
a material with gain is an attractive candidate for several applications. In the quest for a perfect lens
\cite{Pendry2003} the gain is supposed to compensate for the inevitable losses in the 
frequency band where the negative permittivity is achieved. Plasmons, which are considered to be 
ideal candidates for the sub-wavelength manipulation of light, suffer from losses as well.
Here too, combination with a gain medium is supposed to compensate for the losses. Some authors
argue that in this way the surface plasmon amplification by stimulated emission of radiation (SPASER) 
can be achieved, similar to the usual laser \cite{Bergman2001}--\cite{Seidel2005}. While all this is true, 
and our bounds show that real resonances may exist in such media, we can explicitly show that these 
resonances are not necessarily 
the localized lossless plasmons, but may as well be associated with global modes. Consider the 
spatial spectrum corresponding to this case -- see Fig.~2. The upper branch of the essential spectrum is indeed 
approaching zero as with the usual complex plasmon. In a continuously inhomogeneous object there may be essential
spectrum going right through zero in this case. Hence, perfect real plasmons are possible in classical electromagnetics 
(at least mathematically). 
However, in Fig.~2 it is the {\it discrete} eigenvalue, which 
is now the closest to zero, and it has a global eigenfunction associated with it, not a
localized one. Our numerical calculations confirm that the complex plasmon mode and the 
mode of this real resonance indeed look different. Note also that the angular frequency of this resonance 
may in practice coincide with the one of plasmon. 
There is an important difference, though, between the real laser resonances described above and the present 
resonance. If the medium parameters are such that the zero of the complex plane is situated {\it inside} the convex 
envelope of the essential spectrum, then a real {\it quasi-static} resonance can be achieved.
Hence, the mode may be confined to a very small volume, if the object's volume is small.
It may even be enough to reduce the volume of the part with gain only to achieve confinement.

In summary, we have presented a unified approach to macroscopic
electromagnetic resonances in finite inhomogeneous three-dimensional objects. 
We have analyzed the dynamics of the spatial spectrum of the pertaining volume
integral operator as a function of the angular frequency and constitutive parameters,
and were able to recover and generalize all known resonances in this way. In addition, we have 
confirmed the possibility and established conditions for the existence of a real quasi-static resonance 
in media with negative permittivity and gain leading to the volume-dependent light confinement.

\end{document}